\documentstyle[aps,epsf,multicol]{revtex}

\def\p{{\bf p}}
\def\q{{\bf q}}

\def\A{{\bf A}}

\def\be{\begin{eqnarray}}
\def\ee{\end{eqnarray}}
\def\bmat{\begin{pmatrix}}
\def\emat{\end{pmatrix}}

\def\nn{\nonumber}

\def\str{\sigma^{\rm tr}}
\def\Tr{{\,\rm Str\,}}

\begin{document}
\draft
\title{Quantum transport in parallel magnetic fields: A
realization of the Berry-Robnik symmetry phenomenon}
\author{Julia S. Meyer$^1$, Alexander Altland$^1$, and
  B.L. Altshuler$^2$}

\address{$^1$
  Institut f\"ur Theoretische Physik, Universit\"at zu K\"oln, 50937 K\"oln, Germany\\
  $^2$ Physics Department, Princeton University, Princeton,
  NJ 08544, USA\\ and NEC Research Institute, 4 Independence Way,
  Princeton, NJ 08540, USA}
\date{\today}
\maketitle

\begin{abstract}
  We analyze the magnetoconductance of two-dimensional electron and
  hole gases (2DEGs) subject to a {\it parallel} magnetic field.  It
  is shown that, for confining potential wells which are symmetric with
  respect to spatial inversion, a temperature-dependent weak localization signal exists
  even in the presence of a magnetic field. Deviations from this
  symmetry lead to magnetoconductance profiles that contain
  information on both, the geometry of the confining potential and
  characteristics of the disorder.
\end{abstract}

\vspace*{-0.3cm}

\begin{multicols}{2}
\narrowtext

Weak localization (WL) corrections to the conductivity~\cite{WL}
and magnetoresistance of two-dimensional systems in perpendicular
magnetic fields~\cite{mag-res} have been studied extensively for
many years. 
These phenomena originate in the constructive interference of
time-reversed electron trajectories. The magnetic field breaks
time-reversal invariance and, therefore, suppresses the interference.
Considerably less attention has been directed to the effect of an {\it
  in-plane} magnetic field on WL phenomena.  Truly two-dimensional
systems would not feel the orbital effect of an in-plane field at all --
the paths within the plane enclose no flux.  In {\it real} systems,
however, the microscopic profile of the wave functions in the
transverse, or $z$-direction leads to a non-vanishing magnetic
response.  Early works on this phenomenon focused on disordered
metallic films~\cite{AlAr81,DuKh84}, where size quantization is
absent, and two-dimensional electrons subject to {\it short-range}
disorder~\cite{Falko,exp_F}.  A recent paper~\cite{rough-int}
considers systems with rough interfaces, as, e.g.,  Si MOSFETs are
believed to be~\cite{exp_Si}.

In this Letter we analyze the complementary case where the motion of
the carriers in $z$-direction is not completely stochastic.  Such
scenarios are realized, e.g., in a gas of electrons or holes on a
GaAs/AlGaAs interface.  The mobility in these systems is limited by a
{\it long-range} random potential, $V(x,y,z)$, created by charged
impurities located far from the interface.  The
$z$-dependence of this potential is probably weak.  In the
approximation that neglects this dependence, $V =V(x,y)$, the in-plane
motion can be separated from the motion in $z$-direction.  Under these
conditions, WL effects acquire non-universal features, depending on
the structure of the confining potential, $W(z)$.  Thus, monitoring WL
signals one can reveal information on the microscopic structure of the
potential well.  Specifically, the temperature and in-plane magnetic
field dependencies of the conductivity $\sigma(T,H)$ are sensitive to
the symmetry of the confining potential under reflection, ${\cal P}_z:
z\to -z$.  Further, $\sigma(T,H)$ qualitatively depends on the number
of occupied carrier subbands, $M$; The single subband ($M=1$) case
turns out to be special and is characterized by quite unusual
magnetoresistance.

\begin{figure}[h]
\centerline{\epsfxsize=2in\epsfbox{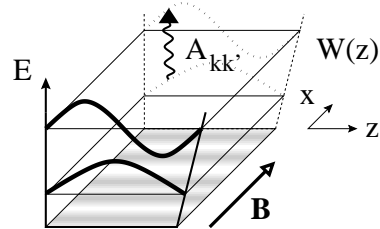}}
\caption{\label{Fig1} Schematic picture of the quantum well.
Two exemplary subband wave functions are shown.
The profile of the
impurity potential is sketched on
the bottom of the well.}
\end{figure}

Let us first discuss the magnetotransport qualitatively.  An
in-plane magnetic field, $H$, manifests itself through the phase
coherence time, $\tau_{\phi}(H)$~\cite{AlAr81,DuKh84}:
 \begin{eqnarray}
\label{old}
\Delta\sigma \!=\! (e^2\!/\pi h)\ln ({\tau/\tau_{\phi}(H)});\quad
1/\tau_\phi(H)\!=\!1/\tau_\phi\!+\!1/\tau_H ,
\end{eqnarray}
where $\Delta\sigma$ is the WL correction to the conductivity,
$\tau$ is the elastic scattering time, and $\tau_H \propto
H^{-2}$.  Consider now the system displayed in Fig.~\ref{Fig1}.
The finite motion in $z$-direction implies a splitting of the
electronic spectrum into different subbands of size quantization.
If $H=0$ and the disorder is $z$-independent, these subbands are
decoupled and contribute separately to the conductivity,
$\sigma$.  Universality of the WL implies that in
this case the correction, $\Delta\sigma$, Eq.~(\ref{old}),
should be multiplied by the
number of the occupied subbands: 
$\Delta\sigma = M (e^2/\pi h) \ln({\tau/\tau_\phi})$.

The magnetic field plays {\it two} complementary roles: it breaks
time-reversal (${\cal T}$) symmetry and (together with the
$z$-dependence of the random potential) couples different
subbands.  In fact, the second role determines the first one: 
${\cal T}$-invariance is
preserved as long as the subbands remain decoupled, since
the vector potential of the parallel
field can be gauged out in each particular subband.
Therefore, the coupling governs the magnetoconductance.  
For strong inter-subband coupling, we return to the disordered film
situation, i.e., Eq.~(\ref{old}) for the WL effect.  When the coupling is weak, 
the WL correction is determined by 
$M$ different decoherence times $\tau^k_H$:
\begin{eqnarray}
\label{new}
\Delta\sigma \!=\! \frac{e^2}{\pi h}{\sum_{k=0}^{M-1}}\!\ln
({\tau/\tau_{\phi}^{k}(H)});\quad
1/\tau_\phi^{k}(H)\!=\!1/\tau_\phi\!+\!1/\tau^{k}_H .
\end{eqnarray}
It turns out that $1/\tau^{k\neq0}_H>0$, i.e., all WL corrections,
except maybe one ($k\!=\!0$), are temperature-independent at $H\neq 0$
and low enough $T$.  Whether $1/\tau^0_H$ vanishes or not depends on
the ${\cal P}_z$-symmetry of the confining potential.  A particularly
interesting situation arises when the system is fully ${\cal
  P}_z$-symmetric: $W(z)\!=\!W(-z)$.  Then, the original Hamiltonian
is invariant under the combination of 
reversal of the magnetic field ($H\!\to\!-H$) 
and ${\cal
  P}_z$-inversions.  This symmetry implies orthogonal rather than
unitary level statistics ~\cite{BeRo86}.  As a result, $1/\tau_H^0=0$,
and $\Delta \sigma\sim \ln ({\tau/\tau_\phi})$ at arbitrary $H$.  (As
$\tau_\phi\propto T^{-p}$, the logarithmic $\sigma (T)$-dependence
persists.)  All other decoherence times $\tau_H^{k\neq 0}$ are
proportional to $H^{-2}$~\cite{fn1}. Accordingly, the WL correction
reads $\Delta\sigma_{\rm s}(H,T)=(e^2/\pi h)[p\ln T+2(M-1)\ln H]$.  By
contrast, any violation of ${\cal P}_z$-symmetry (by either confining
or disorder potentials) suppresses {\it all} WL corrections,
i.e., $\Delta\sigma_{\rm as}(H,T)=2M(e^2/\pi h)\ln H$ for $M\neq 1$.
Therefore, the WL effects sensitively probes the symmetry properties
of the confining (and disorder) potential.  All in all, it is the
interplay of the three factors -- inter-band coupling, ${\cal
  T}$-invariance, and ${\cal P}_z$-invariance that determines the
conductivity, $\sigma (T,H)$.

A special situation arises when just one subband is occupied, $M=1$.
In the absence of high-lying unoccupied bands, the parallel field has
no effect whatsoever -- a one-band system, being structureless in
$z$-direction, cannot accommodate magnetic flux.  Formally, the vector
potential of the field can be removed by a gauge transformation
[cf.~the analysis below].  Thus, ${\cal T}$-breaking at $M=1$ requires
virtual excursions into unoccupied subbands~\cite{Falko}. This fact
substantially reduces the magnetoconductance: If the random potential
is $z$-independent, a residual effect exists, albeit of high order in
the magnetic field, $\tau_{H,M=1} \sim H^{-6}$.  This dependence can
be understood as follows: The matrix elements controlling the
inter-band hopping are proportional to $H$.  This amounts to a hopping
probability $\sim H^2$.  Since the square of the field is ${\cal
  T}$-invariant, the virtual propagation {\it within} the empty bands
must contribute another $H$ and we arrive at $\sim H^3$ for the ${\cal
  T}$-breaking contribution to the self-energy.  Finally, to obtain a
quantum-mechanical intensity, the propagation amplitudes have to be
squared which brings us to $H^6$.  It is essential that the parallel
field performs both ${\cal T}$-breaking {\it and} subband coupling.
Sweeping the subband through the bottom of the
second subband, a crossover $(\tau_{H} \!\sim\!
H^{-2}) \leftrightarrow (\tau_{H} \!\sim\!H^{-6})$ in the WL profile
should be observed.
Table~\ref{tab} displays a summary of the WL signals to be
expected for a given subband population and symmetry configuration.

To derive  these results we start from the 
Hamiltonian 
\begin{eqnarray}
\label{hamiltonian}
    {\cal H}=-\frac1{2m}(\partial+iHz{\bf e}_y)^2+W(z)+V(x,y)\,
\end{eqnarray}
of an electron subject to a confining potential, $W$, and
lateral disorder, $V$.
Later on, we will relax the condition of strict $z$-independence
of $V$.  In the following, we consider only orbital coupling to
the magnetic field; Zeeman splitting and spin-orbit scattering
will be discussed elsewhere. Except for our final results,
$\hbar=c=e=1$.
 
It is convenient to project the Hamiltonian, Eq.~(\ref{hamiltonian}),
onto a basis of eigenfunctions, $\phi_k(z)$,
of the transverse part of the Hamiltonian
$[-\partial_z^2/(2m)+W(z)]$ $\phi_k=\epsilon_k\phi_k$:
\begin{eqnarray}
\label{eq:1}
  {\cal H}_{kk'}=(\frac{-\partial_x^2}{2m}+V(x,y)+
  \epsilon_k)\delta_{kk'}\!-\!\frac1{2m}\big(\partial_y\!-\!i\hat
A\big)^2_{\,kk'}\,,
\end{eqnarray}
where the matrix elements
\begin{eqnarray}
A_{kk'}=-H\int\!dz\,\phi_k(z) z \phi_{k'}(z)\equiv -H d_{kk'}
\end{eqnarray}
measure the degree of ${\cal P}_z$-violation.  If the system is
${\cal P}_z$-symmetric, then $A_{kk}=0$. Also notice that the
explicit structure of the matrix $\hat A$ depends on the choice of gauge. 

To assess transport properties we need to 
evaluate disorder averaged
products of Green functions.
In particular, WL corrections are 
determined by
the two-particle Cooperon propagator.
The key elements of this calculation,
safe for the presence of a subband structure,
are largely standard.
The main result 
can be expressed through the Cooperon matrix, 
$({\cal C}_\q)_{kk'}$:
\begin{eqnarray}
&&\Delta\sigma=-\frac{2}{\pi}{\sum}_{k=0}^{M-1}\int(dq)({\cal
C}_\q)_{kk}\,;\label{cond}\\
({\cal C}_\q^{-1})_{kk'}\!&=&\!\big[(\q\!-\!\!
2A_{kk}{\bf e}_y)^2\!+\!\!\!\sum_{k''=0}^{M-1}\!\!
\frac{{\cal X}_{kk''}H^2}{D_k}\!\big]\delta_{kk'}\!+\!\frac
{{\cal X}_{kk'}H^2}{\!\!\!\sqrt{D_kD_{k'}}};\nonumber\\
{\cal
X}_{kk'}\!&=&\!(1-\delta_{kk'})d_{kk'}^2\frac{D_k\!+\!D_{k'}}
{(\Delta_{kk'}\tau)^2+1},\label{coop}
\end{eqnarray}
where $D_k$ is the diffusion constant of $k$-th subband, and
$\Delta_{kk'}=\epsilon_k-\epsilon_{k'}$.  
The decoherence rate, $1/\tau_\phi$, enters as a lower cut-off for the
$q$-integration.  To understand the meaning of
Eq.~(\ref{coop}), let us compare it with the familiar equation
for the Cooperon in the absence of a subband structure, 
${\cal C}_\q^{-1}\sim q^2+( D\tau_H)^{-1}$.  
If the ${\cal P}_z$-asymmetry of the confining potential is weak, 
i.e., $A_{kk}$ is small,
then 
\begin{eqnarray}
({\cal C}_\q^{-1})_{kk}\sim q^2+\lambda_k; \qquad 1/\tau_H^k=
\alpha_k\lambda_k,
\end{eqnarray}
where
$\lambda_k$ are the eigenvalues of the matrix 
$({\cal C}_{\q=0}^{-1})_{kk'}$ and
$\alpha_k$ is a combination of the diffusion constants
$\{D_0,\,\dots\,,D_{M-1}\}$.

How do the eigenvalues depend on the magnetic field? For $H=0$, all
eigenvalues vanish trivially. Switching on a magnetic field leads to a
coupling of the formerly independent subbands and, thereby, to a set
of $M-1$ positive eigenvalues $\lambda_{k\neq0}$. As a result $M-1$ Cooperon
modes cease to contribute to the low temperature WL signal. However,
the lowest eigenvalue, $\lambda_0$, plays a special role: For a
perfectly symmetric potential, it remains zero implying that a singe
massless Cooperon mode survives application of a magnetic field. This
is a direct manifestation of the Berry-Robnik
phenomenon~\cite{BeRo86}.

For a formal proof note that $A_{kk'}=0$ for even $k+k'$, since ${\cal
  P}_z$-symmetry implies definite and alternating parity of the
eigenstates.  As a result the determinant of the Cooperon matrix,
${\cal C}_{\q=0}$, also vanishes.  Indeed, it is straightforward to
verify that $({\cal C}_{\q=0}^{-1}){\bf X}=0 $, where ${\bf X}$ is a
$M$-component vector

\vspace*{-0.15cm}

\begin{eqnarray}
{\bf X} \equiv {\cal N}^{1/2} \sum_k(-)^k\sqrt{D_k}\,{\bf
  e}_k,\qquad {\cal N}= \Big(\sum_k D_k\Big)^{-1}.\nonumber
\end{eqnarray}

\vspace*{-0.15cm}

Now let ${\cal P}_z$ be slightly violated, either due to asymmetry of
the confining potential or due to the impurity potential.  In this
case the matrix elements $A_{kk'}$, $k+k'$ even, become finite.  To
first order  perturbation theory, the lowest
eigenvalue shifts by $\delta\lambda_0(\q)={\bf X}^T{\cal
  C}_\q^{-1}{\bf X}$.  With Eq.~(\ref{coop}) this evaluates to  $\delta\lambda_0^{\rm (as)}(H) \sim{\cal
  N}[{\sum}_{k+k'\,{\rm even}}{\cal X}_{kk'}+ {\cal
  N}{\sum}_{k,k'}D_kD_{k'} (d_{kk}\!- \!d_{k'k'})^2]H^2$.

Before considering concrete realizations of $W(z)$, let us explore how
an additional weak $z$-dependent disorder potential, $\delta V(x,y,z)$,
affects the Cooperon zero mode. $z$-dependent scattering leads to
additional coupling between the subbands. At sufficiently high
magnetic fields, where the field-induced coupling dominates the
impurity-induced coupling, the lowest eigenvalue can again be
evaluated perturbatively:
\begin{eqnarray}
\delta\lambda_0^{\rm (imp)}\sim{\cal N}\nu
\int\!d^3{\bf r}\,\langle \delta V ^2({\bf r})\rangle
\sum_{k+k'\,{\rm odd}} \int\!dz\,\phi_k^2\phi_{k'}^2\,,
\nonumber
\end{eqnarray}
or $\delta\lambda_0^{\rm (imp)}\sim {\cal N}/\tau'$, where $\tau'$ can
be understood as an inter-subband scattering time.  At small $H$,
however, the dominating coupling mechanism is scattering in
$z$-direction.  In this case the lowest eigenvalue $\lambda_0^{\rm
  (imp)}(H) \sim{\cal N}{\sum}_{k,k'}[{\cal X}_{kk'}+ {\cal
  N}D_kD_{k'} (d_{kk}\!- \!d_{k'k'})^2]H^2$ corresponds to the vector
${\bf X}_l=\sum_k\sqrt{D_k}\,{\bf e}_k$. Thus, $1/\tau_H^0$ increases
as $H^2$ for small $H$ and saturates at $H\sim H_c\sim\Delta/v_{\rm
  F}\sqrt{\tau/\tau'}/d$ (where $\Delta$ is the typical energy
separation between subbands, $v_{\rm F}$ the Fermi velocity, and $d$
the width of the quantum well) if the confining potential is ${\cal
  P}_z$-symmetric.

To illustrate our results on a simple and experimentally relevant example, let us consider a
two-subband system, $M=2$. Assuming for simplicity that 
$D_0=D_1\equiv D$,  the $2\times2$ Cooperon takes the form
\begin{eqnarray}
{\cal C}^{-1}\!=\! {1\over D}\left(\matrix{D (\q\!-\!\A)^2\!+\! {\cal
      X}_{01}H^2\!+\! \frac1{\tau_\phi} \!\!\!\!\!\!\!\!\!\!&  {\cal
      X}_{01}H^2\cr {\cal X}_{01}H^2&\!\!\!\!\!\!\!\!\!\!
    D(\q\!+\!\A)^2\!+\!{\cal X}_{01}H^2\!+\! \frac1{\tau_\phi} }\right),\nn
\end{eqnarray}
where $\A=H(d_{00}-d_{11}){\bf e}_y$, and ${\cal
  X}_{01}$ obtains from Eq.~(\ref{coop}).

At small fields, $H\ll H_\phi\!=\!({\cal X}_{01}\tau_\phi)^{-1/2}$,
the magnetoconductance is insensitive to the ${\cal P}_z$-symmetry and
\begin{eqnarray}
\frac{\sigma(H)-\sigma(0)}{e^2/(2\pi^2\hbar)}\simeq{\cal X}_{01}\tau_\phi
H^2=\frac{2D(e/\hbar)^2d_{01}^2\tau_\phi}{1+(\Delta_{10}\tau/\hbar)^2}H^2,
\end{eqnarray}
independent of the dipole elements $d_{00}$ and $d_{11}$. (Notice that
$\sigma(H)-\sigma(0)$ vanishes in the limit of infinitely separated
bands, $\Delta_{10}\to\infty$, reflecting the behavior of isolated
subbands.)

\begin{figure}[h]
\centerline{\epsfxsize=2.75in\epsfbox{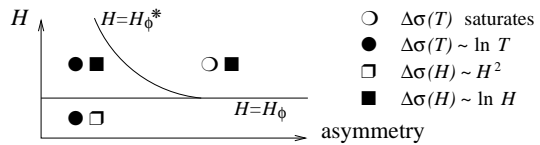}}
\caption{\label{Fig2} Different regimes of $T$- and $H$-dependence.}
\end{figure}

At large magnetic fields, $H\gg H_\phi$, and in the fully symmetric
case, diagonalization of the Cooperon matrix yields
$1/\tau_\phi^0(H)=1/\tau_\phi$ and $1/\tau_\phi^1(H)=2{\cal
  X}_{01}H^2$. While the second term leads to the usual logarithmic
field dependence of $\Delta \sigma$ (cf.~Eq.~(\ref{new})), the field
{\it in}dependence of the second term implies that the conductance
continues to exhibit logarithmic scaling with temperature (through the
$T$-dependence of $\tau_\phi$) at these
large fields.  Slight violation of the symmetry results in a shift of
both eigenvalues
$\delta(1/\tau_\phi(H))=D(e/\hbar)^2(d_{00}-d_{11})^2H^2$. Thus, the
temperature dependence remains as long as
$H<H_\phi^*=\hbar/[e(d_{00}-d_{11})\sqrt{D\tau_\phi}]$. For larger
fields (or stronger asymmetry) the $T$-dependence saturates and the
slope of the $\sigma(\ln H)$-dependence doubles (see discussion after
Eq.~(\ref{new})). The regimes with different parameter dependences of
the conductance are schematically shown in Fig.~\ref{Fig2}.

\vspace*{-0.3cm}

For concreteness, let us list the matrix elements $d_{kk'}$ for two
common realizations of confining potentials: (i)~For a symmetric box potential of
width $d$, $d_{00}=d_{11}=0$ and $d_{01}=-16d/(9\pi^2)$.  Adding a
small perturbation $\delta W(z)=wz$ to the confining potential yields
the diagonal term $d_{00}-d_{11}=4(16d/(9\pi^2))^2w/\Delta_{10}$.
(ii)~For an asymmetric triangular potential well, $W(z)=\infty$ for
$z<0$ and $W(z)=wz$ for $z>0$, one obtains $d_{01}\approx 0.67(2mw)^{-
  1/3} < d_{00}-d_{11}\approx 1.17(2mw)^{- 1/3}$.

\begin{figure}[h]
\centerline{\epsfxsize=2.75in\epsfbox{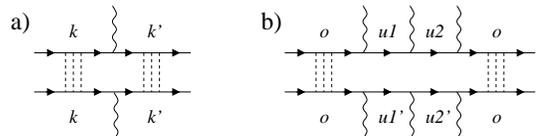}}
\caption{\label{high_virtual} Basic diagrams for a) $M>1$, and b) $M=1$.
The wavy lines show interactions with the magnetic field while the
  dashed lines represent impurity scattering.}
\end{figure}

\vspace*{-0.3cm}

What happens in the case of just one occupied subband? As discussed
above, an in-plane magnetic field does not affect the single
Cooperon mode in the ${\cal P}_z$-symmetric case.  However, for broken
${\cal
  P}_z$-symmetry,
virtual transitions into empty bands lead  to a field and momentum dependent contribution
$\Sigma(\hat A,\p)$ to the self energy of the Green functions of
the {\it occupied} subband. The situation is depicted
schematically in Fig.~\ref{high_virtual}, where the relevant
contribution to the Cooperon
(the two particle Green function) is shown for
$M>1$ (left) and $M=1$ (right).
In the latter case, sixth order scattering off the vector
potential is needed to generate a field dependent contribution.
The self energy can be presented in the form $\Sigma(\p)= {\cal
  D}_{\rm v}(\p)+p_y{\cal A}_{\rm v}(\p)$, where ${\cal D}_{\rm
  v}({\cal A}_{\rm v})$ contains only even (odd) terms in the
magnetic field.
In contrast to ${\cal D}_{\rm v}$ the second term violates
${\cal T}$-invariance by shifting the vector potential:
$A_{00}\to A_{00}+ {\cal A}_{\rm v}(\p)$~\cite{fn3}.
The corresponding magnetic scattering rate
equals
\begin{eqnarray}
\frac{1}{\tau_H}=\frac{D_0}{16}\Big(v_{\rm F}^2\sum_{k,k'>0}\frac{A_{0k}
  (A_{kk'}\!-\!A_{00} \delta_{kk'})A_{k'0}} {\Delta_{0k}\Delta_{0k'}}
\Big)^2\,.\nonumber
\end{eqnarray}
$z$-dependent scattering modifies this result.  The mixing of the
subbands by the disorder, $\delta V({\bf r})$, brings a finite
contribution to $1/\tau_H$ already at second order in the magnetic
field.  Calculation, as performed in Ref.~\cite{Falko}, gives
\begin{eqnarray}
\frac{1}{\tau_H}
\sim\nu v_{\rm F}^2
\int\!d^3{\bf r}\,\langle \delta V^2({\bf r})\rangle
\sum_{k,k'>0}\frac{ A_{0k}A_{k'0}}{\Delta_{k0}\Delta_{k'0} }
\int\!dz\,\phi_0^2\phi_k\phi_{k'}\,.
\nonumber
\end{eqnarray}
This leads to a $H^2 \to H^6$ crossover at
the characteristic field $H_c^{M=1}\sim\sqrt{\Delta/D}(\tau/\tau')^{1/4}/d$.

Finally, let us briefly outline the strategy of the derivation of the above
results.  Although fully perturbative, the intricate interplay of
various scattering mechanisms in the problem suggests to employ the
formalism of functional integration (as an alternative to direct
diagrammatic perturbation theory).  Straightforward adaption of the
standard scheme~\cite{nlsm} of deriving a field theory for disordered
conductors to the structure of the present problem produces a model
with effective action
\begin{eqnarray}
S={\sum_{k=0}^{M-1}} S_0[Q_k]-\frac{\pi\nu}4{\sum_{k,k'=0}^{M-1}}
{\cal X}_{kk'}\Tr\!\Big(\str_3Q_k\str_3Q_{k'}\Big).\nonumber
\end{eqnarray}
Here the matrices $Q_k$ describe diffusive motion in subband $k$,
controlled by the standard (field-dependent) action
$S_0[Q_k]$~\cite{nlsm}. The second
term in the action describes the field-induced coupling between the
subbands.  Notice the structural similarity to the types of actions
appearing in problems with parametric correlations~\cite{parametric}.
Like in those cases, it is the second term that couples the otherwise
un-inhibited degrees of freedom $Q_k$ and, thereby, makes diffusion in
the Cooperon channel massive. Quantitatively, second order
expansion~\cite{nlsm} around the saddle point configurations $Q_k
\equiv \Lambda$ readily produces the Cooperon propagators discussed
above.

We have shown that the magnetoresistance of two-dimensional
electron gases in an in-plane field responds sensitively to both
the geometric structure of the confining potential and the nature
of the impurity scattering.  Those phenomena are intimately
related to the Berry-Robnik symmetry mechanism~\cite{BeRo86}. We
believe that the response in the magnetoconductance profile
should be visible in experiment.

We thank V.I.~Fal'ko for many valuable discussions, and for sending us
Ref.~\cite{FJ01} (which discusses the $H^6$-type
magnetoresponse for the one-subband system in more detail) prior to
publication. We also thank C.M.~Marcus for discussing potential
experimental realizations.


\end{multicols}


\vspace*{-0.25cm}

\widetext

\begin{table}
\begin{center}
\caption{\label{tab} Field-dependent decoherence times, $\tau_H^{(k)}$.
Here $d$ sets the scale for the width of the quantum well, $\Delta$ is the
typical energy separation between subbands, 
$D$ the diffusion constant, $\tau$ the
mean scattering time, $\tau'$ the
mean {\em transverse} scattering time, and $v_{\rm F}$ the Fermi velocity.}
\begin{tabular}{|l||l|l|}
& $M=1$ & $M>1$\\
\hline
\hline
${\cal P}_z$-symmetry & $1/\tau_H=0$ &
$1/\tau_H^0=0$, \enspace $1/\tau_H^{k\neq 0}\sim D/{(\Delta\tau)}^2\,(Hd)^2$\\
\hline
\hline
no ${\cal P}_z$-symmetry due to &&\\
- confining potential, $W(z)\neq W(-z)$ & $1/\tau_H\sim D\,({v_{\rm
F}}/{\Delta}
)^4\,(Hd)^6$ & $1/\tau_H^k\sim D/{(\Delta\tau)}^2\,(Hd)^2$\\
\hline
- disorder, $V=V(x,y,z)$ & $1/\tau_H\sim (v_{\rm
F}/{\Delta})^2/\tau'\;(Hd)^2$ & $1/\tau_H^0\sim
\min\{D/{(\Delta\tau)}^2\,(Hd)^2,1/\tau'\}$,\\
&& $1/\tau_H^{k\neq0}\sim D/{(\Delta\tau)}^2\,(Hd)^2$\\
\end{tabular}
\end{center}
\end{table}



\begin{references}
\bibitem{fn1} The difference to the perpendicular case, where
  $\tau_{H_\perp}\sim H_\perp^{-1}$, follows from the fact that
  the characteristic area threaded by one flux quantum, $d
  l_{H_\parallel} \sim H_\parallel^{-1}$, is linearly related to
  the fixed width $d$ of the potential well.
\bibitem{fn3} In principle, the diamagnetic term, $A^{\rm
    dia}_{\;kk'}=(A^2)_{kk'}$, has to be taken into account.  It
  turns out, however, that -- to lowest order in $H$ -- the
  Cooperon mass is determined solely by the vector potential
  $A_{kk'}$. Note that in the symmetric case ${\cal A}_{\rm
    v}(\p)=0$ and, thus, $\tau_H^{-1}$ vanishes as expected.
\end{references}
\end{document}